# AI-based software for lung nodule detection in chest X-rays - Time for a second reader approach?

Ohlmann-Knafo S[1], Ramanauskas N[2,3], Huettinger S[1], Jeyakumar JE[1], Barušauskas D[2], Bielskienė N[2], Naujalis V[2], Bialopetravičius J[2], Ražanskas J[2,3], Samuilis A[3], Dementavičienė J[3], Pickuth D[1]

[1]*CaritasKlinikum Saarbruecken, Diagnostic and Interventional Radiology, Rheinstrasse 2, 66113 Saarbruecken, Germany*

[2]*Oxipit, Saulėtekio al. 15, LT-10224 Vilnius, Lithuania*

[3]*Department of Radiology, Nuclear Medicine and Medical Physics, Institute of Biomedical Sciences, Faculty of Medicine, Vilnius University, LT-08661 Vilnius, Lithuania*

# Abstract

**Objectives**

To compare artificial intelligence (AI) as a second reader in detecting lung nodules on chest X-rays (CXR) versus radiologists of two binational institutions, and to evaluate AI performance when using two different modes: automated versus assisted (additional remote radiologist review).

**Methods**

The CXR public database (n = 247) of the Japanese Society of Radiological Technology with various types and sizes of lung nodules was analyzed. Eight radiologists evaluated the CXR images with regard to the presence of lung nodules and nodule conspicuity. After radiologist review, the AI software processed and flagged the CXR with the highest probability of missed nodules. The calculated accuracy metrics were the area under the curve (AUC), sensitivity, specificity, F1 score, false negative case number (FN), and the effect of different AI modes (automated/assisted) on the accuracy of nodule detection.

**Results**

For radiologists, the average AUC value was 0.77±0.07, while the average FN was 52.63±17.53 (all studies) and 32±11.59 (studies containing a nodule of malignant etiology = 32% rate of missed malignant nodules). Both AI modes - automated and assisted, produced an average increase in sensitivity (by 14% and 12%) and of F1-score (5% and 6%) and a decrease in specificity (by 10% and 3%, respectively).

**Conclusions**

Both AI modes flagged the pulmonary nodules missed by radiologists in a significant number of cases. AI as a 'second reader' has a high potential to improve diagnostic accuracy and radiology workflow. AI might detect certain pulmonary nodules earlier than radiologists, with a potentially significant impact on patient outcomes.

**Key points**

- *AI software improves lung nodule detection rate compared to radiologist evaluation. The most significant improvements were measured for the most subtle pulmonary nodules.*

- *AI software as a second reader versus the more known primary reader approach has the advantage of reducing the irrelevant AI suggestions shown to the radiologist. Especially, the AI assisted mode (additional reviewing remote radiologist) might mitigate the risk of radiologist desensitization.*

- *AI as a second reader can enhance radiologists' workflow and even be helpful for less experienced readers.*

**Abbreviations**

AI    Artificial Intelligence

AUC  Area under the curve

CAD  Computer Assisted Diagnosis

CT    Computed tomography

CXR  Chest X-ray Imaging

FN    False negative

ROC  Receiver operating characteristic

# Introduction

Chest X-ray imaging (CXR) is one of the most frequently requested radiological examinations worldwide and is the modality of choice for primary chest evaluation [1]. This is because CXR enables a fast diagnostic clarification of thoracic symptoms to a high degree combined with lower radiation exposure and lower costs compared to higher-resolution computed tomography (CT). However, CXR interpretation may be challenging, especially in view of the increase in the radiology workload, radiological staff shortage, and the risks of delayed diagnoses and medical malpractice lawsuits [2, 3, 4, 5].

There is no established consensus on the general error rate of undetected radiological findings, and its clinical significance is not always clear [6, 7]. It is claimed that up to 30% of all abnormal CXRs contain at least one pathology that is not correctly identified [7]. For specific radiological findings, such as pulmonary nodules, the false negative rate varies from 19% to 90% in older publications [8-12]. The dataset used in our study originates from a publication where the average area under the curve (AUC) value for pulmonary nodule detection was 83% [13]. Notably, even digital radiographic technologies could not resolve the issue of undetected or misinterpreted lung nodules [14, 15].

Different studies have concluded that population-wide CXR screening does not reduce the mortality of patients with lung cancer [16]. Screening of high-risk patients with CT can reduce lung cancer mortality to a certain extent, but various issues remain, such as unnecessary tests and incidental CT findings [17, 18]. Furthermore, both the start and progress of lung cancer screening programs in Europe have been slow and laborious [19].

However, apart from the screening programs, scientific data are not sufficient to base the claim that there is no clinical value in the early detection of malignant pulmonary nodules on CXR, especially when it comes to individual patients.

Therefore, a general goal would be to improve lung nodule detectability on CXR, especially when lung nodules are subtle or hidden by over- or underlying anatomical structures. With better nodule detectability or exclusion on primary CXR, indications for lung CT might be more accurate and unnecessary radiation exposure could be avoided.

For radiological findings other than pulmonary nodules, to our knowledge, there are no studies which evaluated the clinical effect of missed findings on patient outcomes. However, it could be assumed that for findings such as tension pneumothorax or device malpositions it is self-evident that delayed diagnosis can worsen patients' condition [20, 21].

Commercial artificial intelligence (AI) software solutions aim to increase the diagnostic accuracy of CXR. Many are certified for clinical use and are utilized in medical institutions worldwide [22]. Most of them operate as a first reader or a computer-assisted diagnosis (CAD) device and provide preliminary suggestions for the reporting radiologist during CXR interpretation. Although various studies on AI solutions have shown improvements in sensitivity for detecting various findings, such as pulmonary nodules [14, 23], consolidations [22], or pneumothoraces [1,22], AI as a first reader usually produces a significant decrease in specificity, which in a clinical setting implies a large number of false-positive AI suggestions [24].

In a controlled study setting, the increase in sensitivity and decrease in specificity suggest a significant net positive effect on overall accuracy. In a real-time clinical setting, however, the effect of numerous false-positive AI suggestions is difficult to measure, and frequent false-positive AI suggestions could disincentivize the radiologist to pay attention to the AI feedback, resulting in no quantifiable value in clinical work.

A possible solution to partially mitigate the excessive false-positive AI suggestions and improve radiologists' CXR readings is to utilize AI as a second reader. After the radiologist has interpreted the CXR and wrote the report, AI as a second reader analyzes and compares the output of the CXR and radiologist's report and highlights the CXR with a potentially missed finding for quality assurance purposes. A different and further approach would be to redirect the AI-flagged CXR to an additional remote human reader.

In this study, AI software (ChestEye Quality; Oxipit, Vilnius, Lithuania) was used as a second reader in two operation modes: automated and assisted (the last with an additional remote radiologist).

Our aim was to identify possible improvements in the accuracy of detecting pulmonary nodules on CXR when using both AI operation modes and to compare the results between these two modes.

## Materials and methods

This is a retrospective study on a public dataset.

**The evaluated AI software solution**

The AI software ChestEye Quality, developed by Oxipit, is used as a quality assurance tool that analyzes both CXRs and radiologist reports. The software detects and compares CXR findings with the radiologists' reports. In this context, the CXR with the largest potential for a missed radiological finding is used for a secondary review. This AI system has two alternative modes: automated and assisted.

- Automated mode: All CXR flagged by the AI were subsequently forwarded to the radiologist. The radiologist reviewed the highlighted areas and finalized the report.
- Assisted mode: The CXR automatically flagged by the AI solution is subsequently reviewed by an additional remote reviewer to reduce the number of false positive cases before being forwarded to the radiologist.

This study aimed to measure the effect of both AI operation modes on the accuracy of nodule detection.

**Dataset description**

This study used the Japanese Society of Radiological Technology dataset [9], which was originally published in 2000. The dataset contains 247 CXR and includes positive cases, including pulmonary nodules of various etiologies and with biopsy-level ground-truth information, as well as negative studies without nodular opacities. CXRs containing pulmonary nodules are classified into five different levels of subtlety according to radiologist consensus. The full description of the dataset can be found in the original publication. Histopathological ground truth and subtlety classifications were used for performance evaluations.

**Evaluation of CXR by radiologists**

Eight readers with 3 to 16 years' experience (four radiologists and four radiology residents) from two institutions (Germany and Lithuania) participated in this study. All images were presented in randomized order for the detection of lung nodules. The readers were asked to document the following image information: 1) binary classification representing whether a nodule was visible, and 2) their confidence level about a nodule being visible in the image.

**Processing of the results by AI**

After analysis by each radiologist, the results (CXR and binary labels by the radiologist) were processed by AI. The AI software produced binary labels representing which CXR would be flagged for a second reading, based on the predictions by AI and the data extracted from the radiologist labels.

a. For the automated AI mode, all flagged CXRs were used for further performance analysis.
b. For the AI-assisted mode, all the AI-flagged CXR were additionally reviewed by a remote human radiologist who was asked to exclude any of the flagged CXR by AI, which did not contain any suspicious nodular opacities.

## Results

**Evaluation of the standalone reader performance versus both AI modes**

The metrics used for performance evaluation were the AUC, sensitivity, specificity, F1 score, and number of false-negatives (FN). The metrics of each radiologist were averaged to represent the average reader performance in the dataset (Table 1). The receiver operating characteristic (ROC) curve analysis results are shown in Fig. 1.

| | AUC | Sensitivity | Specificity | F1 score | FN | Malignant FN |
|---|---|---|---|---|---|---|
| Reader 1 (radiology resident) | 0.85 | 0.80 | 0.76 | 0.79 | 31 | 20 |
| Reader 2 (radiologist) | 0.65 | 0.47 | 0.80 | 0.59 | 82 | 53 |
| Reader 3 (radiology resident) | 0.82 | 0.75 | 0.75 | 0.76 | 38 | 23 |
| Reader 4 (radiologist) | 0.77 | 0.73 | 0.47 | 0.63 | 42 | 27 |
| Reader 5 (radiology resident) | 0.72 | 0.64 | 0.72 | 0.67 | 56 | 33 |
| Reader 6 (radiologist) | 0.82 | 0.68 | 0.88 | 0.76 | 49 | 28 |
| Reader 7 (radiologist) | 0.79 | 0.68 | 0.82 | 0.74 | 49 | 26 |
| Reader 8 (radiology resident) | 0.74 | 0.52 | 0.89 | 0.66 | 74 | 46 |
| Avg ± st.dev | 0.77±0.07 | 0.66±0.11 | 0.76±0.13 | 0.70±0.07 | 52.63±17.53 | 32.00±11.59 |

Table 1. Standalone radiologist performance metrics

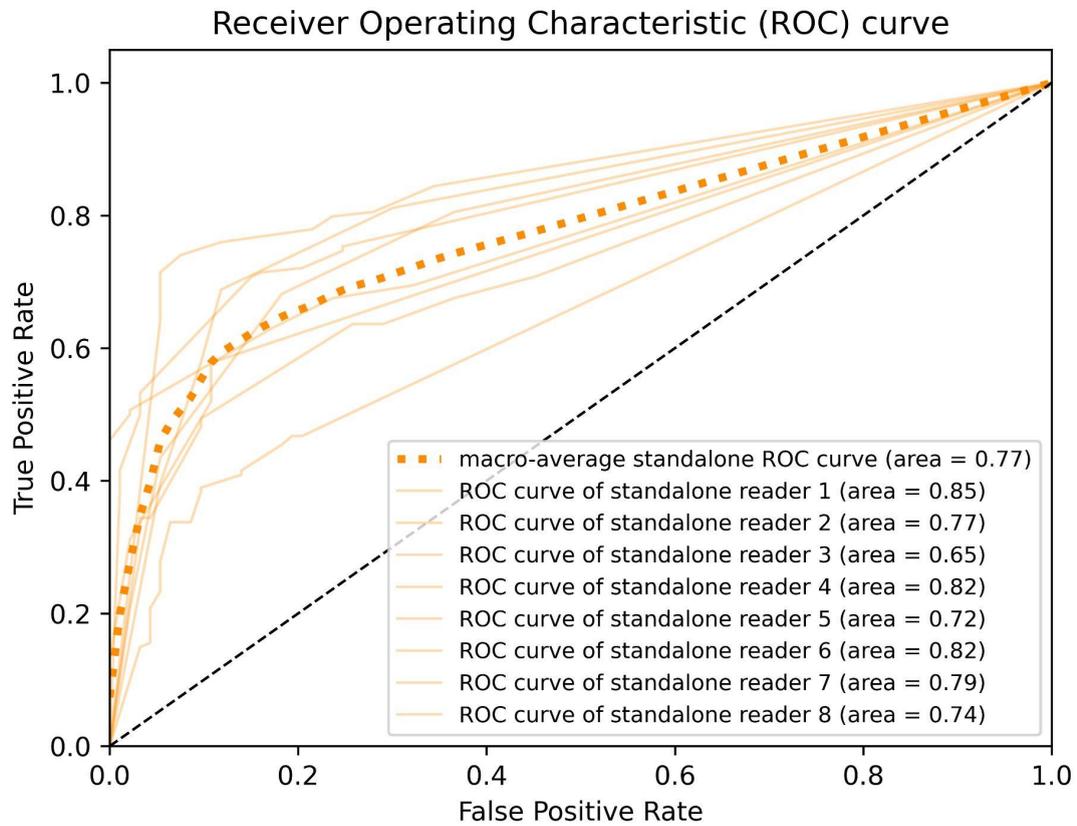

Figure 1. ROC curves for each individual reader

There was significant variation in the accuracy metrics between the radiologists. The AUC values for different readers ranged from 0.65 to 0.85 with an average of 0.77±0.07. The average number of false negative cases was 52.63±17.53 for all CXRs and 32±11.59 for CXRs containing a nodule of malignant etiology. This represents an average rate of missed malignant nodules of 32%.

The reader performance metrics were measured for each nodule subtlety level (from 1 = hardest to 5 = easiest). ROC curves showed distinguishable differences between the average AUC values for different levels of subtlety (Figure 2).

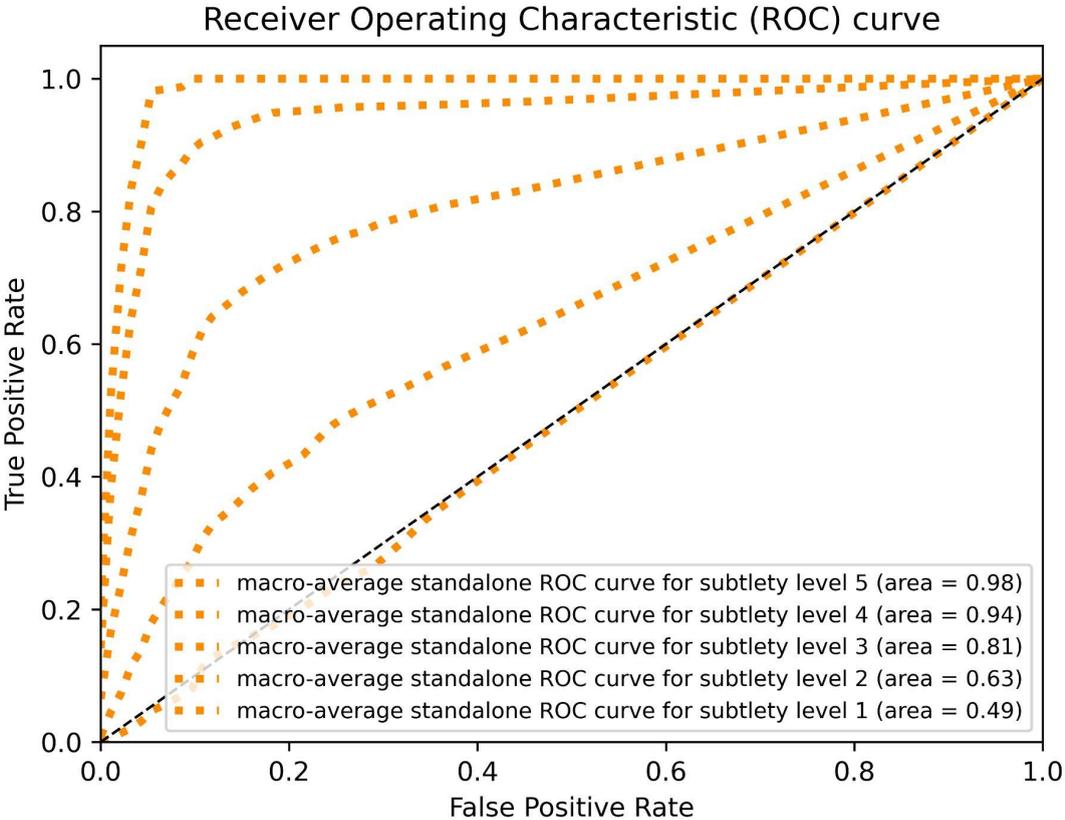

Figure 2. ROC curves for different subtlety levels.

After the evaluation of the standalone radiologist's performance, the effect of AI was measured on the previously selected metrics using both the AI automated and assisted modes.

The AI automated mode processed the images as well as the radiologist's annotations. It then automatically generated a list of studies based on AI suggestions, as well as the analysis of the radiologist annotations, which contained potentially missed pulmonary nodules.

In the AI-assisted mode, the AI-flagged CXR was subsequently reviewed by a third reader - a remote radiologist who was asked to either accept or reject the AI-flagged CXR findings. Subsequently, only accepted cases were sent for a second review to the institutional radiologist. The results are presented in Table 2.

|  | Standalone reader | with AI (automated mode) | with AI (assisted mode) |
| --- | --- | --- | --- |
| Sensitivity | 0.66±0.1 | 0.8±0.04 | 0.78±0.05 |
| Specificity | 0.76±0.13 | 0.66±0.12 | 0.73±0.13 |
| F1 score | 0.7±0.06 | 0.75±0.05 | 0.76±0.05 |
| Total FN | 52.63±15.24 | 30.38±5.89 | 33.38±7.06 |
| Malignant FN | 32±10.54 | 18.75±4.47 | 19.88±5.04 |

Table 2. Effect of applying AI as a quality assurance tool for reader performance metrics

The AI automated and assisted modes produced an average increase of 14% and 12% in sensitivity and a decrease of 10% and 3 % in specificity, respectively. The F1 score showed an average increase of 5% and 6%, respectively, suggesting an overall positive effect of both AI modes.

Furthermore, the number of FN CXRs produced by radiologists was flagged as suspicious by AI automated and AI-assisted mode to be double-checked for a suspected missed nodule. The results are presented in Tables 3 and 4.

| | FN | FN with AI | Malignant FN | Malignant FN with AI |
|---|---|---|---|---|
| Reader 1 | 31 | 22 | 20 | 14 |
| Reader 2 | 82 | 41 | 53 | 26 |
| Reader 3 | 38 | 28 | 23 | 17 |
| Reader 4 | 42 | 28 | 27 | 18 |
| Reader 5 | 56 | 34 | 33 | 20 |
| Reader 6 | 49 | 29 | 28 | 18 |
| Reader 7 | 49 | 23 | 26 | 12 |
| Reader 8 | 74 | 38 | 46 | 25 |
| Avg ± st.dev. | 52.63±17.53 | 30.38±6.78 | 32±11.59 | 18.75±4.86 |

Table 3. Effect of false negative reduction after applying automated suggestions produced by AI (automated mode)

| | FN | FN with reviewed AI | Malignant FN | Malignant FN with reviewed AI |
|---|---|---|---|---|
| Reader 1 | 31 | 25 | 20 | 15 |
| Reader 2 | 82 | 46 | 53 | 29 |
| Reader 3 | 38 | 30 | 23 | 18 |
| Reader 4 | 42 | 28 | 27 | 17 |
| Reader 5 | 56 | 35 | 33 | 20 |
| Reader 6 | 49 | 32 | 28 | 19 |
| Reader 7 | 49 | 27 | 26 | 14 |
| Reader 8 | 74 | 44 | 46 | 27 |
| Avg ± st.dev. | 52.63±17.53 | 33.38±7.82 | 32±11.59 | 19.88±5.41 |

Table 4. Effect of false negative reduction after applying automated suggestions coming from AI and confirmed by the reviewing radiologist (assisted mode)

On average, AI has flagged for second reading 42% (automated) and 37% (assisted) of all FNs and 41% (automated) and 38 % (assisted) of all the FNs containing malignant nodules.

**Analysis of different subtlety levels**

The effect on sensitivity for both AI modes for each of the five nodule subtlety levels was measured (Table 5).

|  | Human reader | Automated mode | Assisted mode |
|---|---|---|---|
| level 5 | 1 | 1 | 1 |
| level 4 | 0.95 | 0.99 (+4%) | 0.99 (+4%) |
| level 3 | 0.73 | 0.85 (+12%) | 0.83 (+10%) |
| level 2 | 0.44 | 0.76 (+32%) | 0.7 (+26%) |
| level 1 | 0.17 | 0.39 (+22%) | 0.36 (+19%) |

Table 5. Sensitivity uplift after AI and reviewing radiologist suggestions for different subtlety levels

There was a significant improvement in the sensitivity values for subtlety levels 1-4. The most significant improvements in the average sensitivity were measured for Level 2 and Level 1 nodules (most subtle). As the sensitivity for level 5 nodules (most obvious) was 100%, meaning every nodule was detected by radiologists, there was no effect on sensitivity for level 5 nodules. A larger increase in sensitivity was measured for the automated AI mode, although the differences between the increases in sensitivity for the two AI modes were marginal.

**Results comparison of AI automated versus AI assisted mode**

Both AI automated and assisted modes demonstrated a significant increase in performance metrics, as indicated by a significant improvement in F1 scores (see Table 2). The automated mode produced a marginally larger increase in average sensitivity but led to a larger reduction in specificity. The assisted mode demonstrated a similar increase in sensitivity with a minor reduction in specificity. Both modes demonstrated a significant decrease in the standard deviation between different radiologists, indicating a significant effect on the performance of radiologists with different individual skill levels. Both modes demonstrated a significant reduction in both malignant and benign FNs.

**Conclusions**

There were large and statistically significant differences in pulmonary nodule detection in CXR between different readers. With increasing subtlety of the pulmonary nodules, there was a significant decrease in the accuracy of the readers. For the subtlest pulmonary nodules (levels 1 and 2), some of which contained malignant nodules, less than half of the nodules were correctly identified. This suggests that there is great potential for the application of different measures for accuracy improvement, as well as uniformizing the performance of readers with different skill levels.

The application of AI as a second reader also contributes to a decrease in specificity, which in clinical practice can result in a significant number of false positives, thus desensitizing the reporting radiologist to the output of the AI software. This can be mitigated by introducing a third reader (a remote radiologist) into the operation pipeline (assisted mode), which can significantly reduce the number of false positives while retaining a comparable improvement in sensitivity.

# Discussion

**Impact of AI as a second reader to the nodule detection accuracy**

Our findings indicate that utilizing AI as a second reader can significantly contribute to the accuracy of pulmonary nodule detection in CXR. There was a statistically significant increase in sensitivity for all radiologists. The most significant improvements were measured for the most subtle pulmonary nodules, a sample that, if detected correctly and on time, can contribute to better patient triage for chest CT, earlier diagnosis of pulmonary cancer/metastases, and potentially better patient outcomes.

**Effect on the uniformization of the performance of readers with different individual skill levels**

Despite the pronounced differences in the individual performance of different readers, the results suggest that utilizing AI as a second reader can significantly reduce the variability in the performance of readers with different individual skill levels. In the real-world setting, this can compensate for the lack of experience or other factors contributing to lower performance, such as reader fatigue or distractions. AI as a second reader could also significantly contribute to a higher standard of care in radiology service providers, where the general level of experience is lower than that in tertiary care or radiology centers, such as primary clinics, screening centers, or institutions with non-radiologists (e.g., radiographers or physicians).

**Benefits of AI as a second reader approach**

Many currently available AI systems function as a first reader approach. When the AI suggestions are provided to the reporting radiologist during the interpretation of the image, AI as a second reader mode provides significant benefits, since it greatly reduces the number of non-useful suggestions to the reader. This is because CXRs are not displayed to the radiologist when there is an agreement between the radiologist and AI. In addition, it allows for using data from the radiologist report/referral, which can be used to reduce the number of FNs. For example, post-biopsy studies that would be highlighted as nodule studies in AI as the first reader scenario are not flagged in a second reader approach by utilizing the data included in the radiologist's report.

The second reader approach also has the potential to mitigate the risks of AI utilization identified in a publication by Gaube et al. [25]. The study identified that there is a significant risk of reducing the overall performance by providing preliminary suggestions to the readers, especially if the recipients of the AI suggestions are non-radiologists, such as physicians with less expertise. The application of AI as a second reader handles the risk of introducing FNs, which would otherwise not be present, and the implementation of a reviewing radiologist in the pipeline (assisted mode) can mitigate the potential negative effect on overall sensitivity.

**Comparison of the two different modes of operation**

Both operation modes (automated and assisted) resulted in increased sensitivity and F1 scores for detecting pulmonary nodules for all readers. Although the assisted mode showed a slightly smaller increase in sensitivity, it also showed a significantly lower decrease in specificity, which contributed to the higher overall F1 score. In a real-time clinical scenario, a large frequency of false positive output for the AI solution can add additional demands to the already strenuous radiologists' workload and disincentivize the reader to the suggestions, thus also minimizing the potential improvement in sensitivity. It is likely that on a dataset representative of a clinical setting, containing a larger number of studies containing various pathologies besides pulmonary nodules, the rate of false positive studies flagged by the AI automated mode would be even more prominent, resulting in a more significant contribution of AI-assisted mode in mitigating the risks of desensitization of the radiologist.

The results of this study indicate that the assisted operation mode can significantly improve the performance in detecting pulmonary nodules on CXR while minimizing the number of false-positive CXRs, resulting in the best of both worlds scenario - improvement in sensitivity without significant sacrifices in specificity.

**Limitations**

There were several significant limitations to this study:

1. The effect of the AI solution as a second reader was measured with the assumption that all FNs were perceptual and not interpretation errors [26]. It was not measured whether the reader reconsidered his diagnosis after the nodules were presented by the AI solution after the initial reading. It is likely that for a fraction of the FN studies where the reader did notice the nodular opacity but considered it unlikely to be a real underlying nodule – an interpretation error or false decision scenario – the radiologist's report would not be readjusted. However, according to different studies, perceptual errors comprise 60 – 90% of the radiological errors [26, 27]. This indicates that the positive effect on sensitivity would likely be retained even in a scenario where this factor would be accounted for.
2. The dataset used was not representative of CXR studies in a real clinical setting. The dataset contains an overrepresented sample of CXR with pulmonary nodules and an underrepresented sample of studies with radiological findings other than pulmonary nodules. In a general sample, the rate of CXRs containing a solitary pulmonary nodule is only around 0.1-0.2% [28]. This is likely to result in underestimation of the effect on the specificity values; however, it should not influence the sensitivity values.
3. The dataset is publicly available from the Internet. This means that the presented CXR did not contain the full pixel depth when displayed on a radiologist's PACS workstation and that several helping diagnostic tools, utilizing the full DICOM image data, were not available. In this sense, radiologist reports might have been more accurate when CXR was displayed with the same image quality and with diagnostic working tools as carried out on radiologists' own workstations.